# A model-driven approach for processing complex events

István Dávid
Department of Measurement and Information Systems
Budapest University of Technology and Economics
Budapest, Hungary
davidi@inf.mit.bme.hu

*Abstract* — By adequate employing of *complex event processing (CEP)*, valuable information can be extracted from the underlying complex system and used in controlling and decision situations. An example application area is management of IT systems for maintaining required dependability attributes of services based on the infrastructure.

In practice, one usually faces the problem of the vast number of distributed event sources, which makes depicting complex event patterns a non-trivial task.

In this paper, I present a novel, model-driven approach to define complex event patterns and directly generate event processing configuration for an open source CEP engine widely used in the industry.

One of the key results of my research work is a textual modeling language called Complex Event Description Language (CEDL), which will be presented by its algebraic semantics and some typical examples.

*D.2.13 [Software Engineering]: Reusable Software – Domain engineering*

I. INTRODUCTION

*A. Motivation*

As information systems are evolving and growing, they become more complex as well. Nowadays, homogenous systems are obsoleted by distributed solutions like service-oriented architecture and the strongly related event-driven architecture. These kinds of systems can take advantage of complex event processing quite successfully. [1][2]

Despite the broad spectrum of possible application areas of complex event processing, the technical solutions are still in the early phase of evolution.

Defining relevant patterns requires precision. Even in the most basic case, one has to specify the sources of events; the metrics, which the event recognition should be based on; and the values of these metrics, which are "relevant", i.e. after the evaluation of the related characteristics of the (complex) event, the results fall in a previously given acceptance domain. It is easy to see, there are *many parameters, not necessarily known* in development time or that may vary as later applied measurements are getting fed back. It is also *not clear, which patterns are "relevant"* and how (what kind of metrics with) they can be measured. The answer can be found in the standards of the given domain – since these disciplines define the high-level requirements against the given system. (E.g.: the requirements of the systems in the IT domain, specified in the standards of COBIT. [3])

Current industrial tools are shipped with a custom language, designed for the given platform and utilize the special features of it; consequently, the language needs to be low-level and *falls short to capture the relatively high-level phenomenon*. (E.g.: the term "critical service" is too general in this form to capture with a platform-specific language.)

As pointed out earlier, it is a common case one having to deal with hundreds of event sources. In practice, the definition of event patterns is an ad-hoc process. As a result, *the consistency of the defined event model cannot be assured.* (E.g. patterns may interfere with each other.) Furthermore, the conformance to standards can be circumstantial.

*B. Goals*

In this paper, a novel, model-driven approach to defining complex event patterns will be presented. I will describe a modeling language (*CEDL – Complex Event Description Language*), which, in part, can be produced from the general-purpose knowledge bases of the given domain. This approach, on the one hand gives a generic and domain-agnostic modeling toolset, yet on the other one, lets the language be easy-to-apply for different domains.

A possible implementation of the semantic knowledge base will be presented and the way how some specific parts of the modeling language can be derived from it.

In order to be formally verifiable, the algebraic semantics of the language had been formalized. I will discuss this algebra in Section III.C and compare to other languages.

II. THE APPROACH

Figure 1. shows the schematics of the approach briefly discussed in this paper.

Modeling of complex events is accomplished by a *textual modeling language*, called CEDL (Complex Event Description Language), which is based on the *General metamodel of the language*. This metamodel defines a set of terms, such as *Event, ComplexEvent, Source*, etc. These concepts are domain-agnostic, i.e. have the same semantics over different domains. Prior to actually defining an event pattern, the general metamodel must be extended by the semantic knowledge of the *Targeted domain* as well as by the *Structural data* of the domain (the concrete event sources), which will be referable in the patterns of the Event-model. Structural data serves as a *Configuration* for the language, already extended by the *Semantic knowledge base*.

*Domain parameters* refer to a set of values taken from historical data, which can affect the business logic, as mentioned in Section I. This data can be collected by monitoring the functioning system. Of course, intelligent analysis is needed prior to the data being fed back.

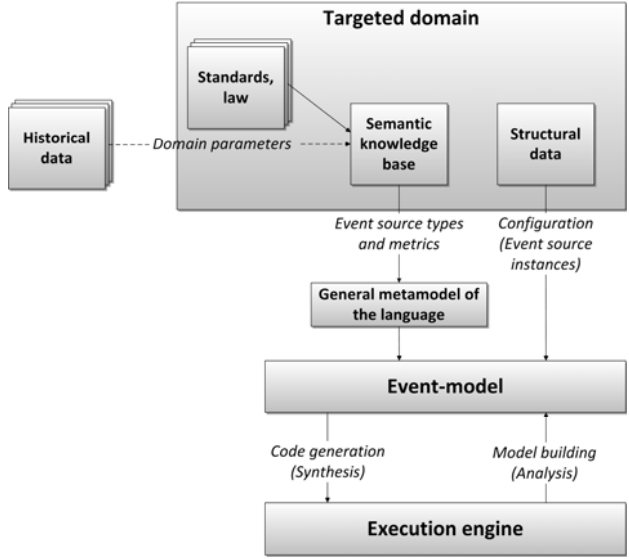

Figure 1. Schematic structure of the system

Finally, source code is going to be generated from the event model for an appropriate *Execution engine*[1]; the code generation is executed automatically by the integrated modeling environment I have implemented using up-to-date Eclipse technologies.

**Example 1** To understand, how the modeling language is assembled by a general and a domain-specific part, consider the following example, taken from the domain of IT infrastructure management. This event pattern describes a case *CPU load* of the resource *Server1* being *at least 90%*.

```
Event CPULoadCritical {
  source Server1
  PercentageMeasurement CPULoad Minimum 90
}
```
□

Here, the emphasized keywords (*Event, source, PercentageMeasurement, Minimum*) are the part of the General metamodel of the language. Meanwhile event source *Server1*, measurement *CPULoad* and the value of the relevant case of the measurement, *90* are defined by the domain (e.g. the underlying system, the feasible measurements, high-level goals, such as SLA's, for example.) Hence we consider these elements as *domain information* concepts.

The support for connecting domain information to general event patterns will be discussed in Section IV.

**Example 2** In the previous example we defined an *atomic* event: it originates from a single source and depicts a simple phenomenon associated with the corresponding source. In contrast, the following one is a *complex event* pattern, which depicts the case of two atomic events being concurrently observable for at least 30 seconds.

---

[1] In my case this was the open-source Java-based engine ESPER. [4]

```
ComplexEvent CriticalServer{
  CONCURRENT_T(CPULoadCritical BackupProblem;
                                    T:Minimum 30)
}
```
□

### III. LANGUAGE ESSENTIALS

As mentioned earlier, the Complex Event Description Language is a textual modeling tool and it is based on a general metamodel, which can be extended by (or customized for) the targeted domain.

#### A. General metamodel

The general metamodel (Figure 2.) consists of two fundamentally different sets of types: the ones which are completely domain-independent and thus of full value (denoted with grey); and the ones which provide interfaces for the domain information to be mapped onto them (denoted with white). (E.g.: the type *Event* is referred in the very same form in the language, but the type *Source* first needs to be configured by structural data of the domain.)

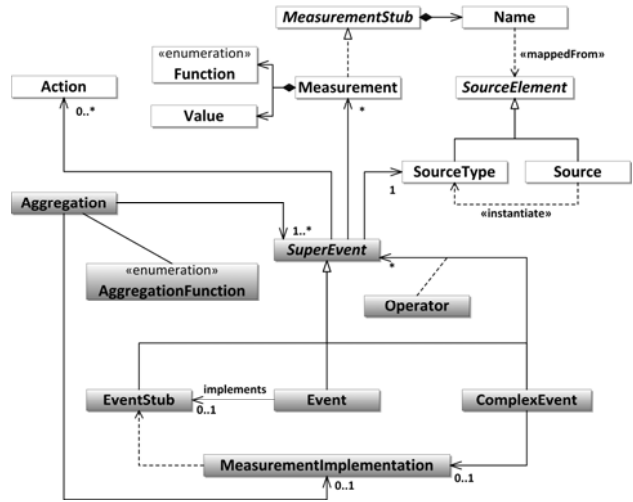

Figure 2. The general metamodel of the language.

#### B. The Event stub design pattern

Design patterns provide general, reusable solutions for commonly occurring problems. Hereby, just the most important one of the CEDL will be defined, the *Event stub*.

The pattern is associated with three, fundamentally different problems. (1) Multiple events logically extend another one (formally an ancestor) by specifying new attributes or overriding the original ones. (E.g.: measurements.) Since every extending event modifies the set of the original attributes, we would prefer to define the measurements in the ancestor without assigning any values, i.e. by defining only *measurement stubs*. (See Figure 2.) (2) The format of the events is known just partially, or a set of events share a common part. (3) An event pattern should be defined for not a concrete event source, just for the type of it; as a result, a single pattern can be later applied for all of the instances of the type.

The solution is motivated by the object-oriented paradigms. Problem (2) requires an abstract base class to be defined, while problem (3) can be solved by using an interface. Problem (1) is a hybrid in this sense: some kind of a base-event needs to be defined, but being able not to assign values to the measurements is a requirement too.

CEDL defines a structure, designed especially for this problem, the *EventStub*. By the formal definition the event stub is a result of an arbitrary number of truncating an event, where truncating means one of the following. (a) *Structural truncating*: Structural elements are getting removed from the event: attributes, action definitions or measurements. (b) *Value truncating*: An assigned value of a structural element is getting removed from the event, thus the remains only indicate the presence of the structural element; the event implementing the event stub is forced to assign a value to the structural element. (In the case of the measurements, we say, the event implements the measurement; thus the element **@ImplementMeasurement** in the language. See Figure 2. : *MeasurementImplementation*.)

By definition, an event stub can refer only to a type of source (*SourceType*), but not to a concrete instance. The implementing event defines the concrete source instance of the type defined in the event stub.

As a result, unnecessary redundancy can be avoided on the level of the model. This design pattern also offers the possibility of defining event patterns associated with only a type of source and later instantiate it. (Keyword **OFTYPE** is used to implement the event stub for every occurrence of sources of the given type.)

**Example 3** This example shows how the Event stub design pattern can be utilized in some typical use cases.
```
EventStub CriticalCPULoadOnWebServer {
  sourceType WebServer
  PercentageMeasurement CPULoad
  actions{
      Action of Type sendWarning
  }
}
```
Accordingly to the previously discussed, the event stub specifies a source type (*WebServer*), a measurement but without assigning any value to it (*CPULoad*) and an action, which can be overridden later.

Here, a solution only for problem (3) will be shown, by an event covering all the instances of a given source type.
```
Event SuspiciousCPULoadOnAnyWS implements
CriticalCPULoadOnWebServer {
  source OFTYPE WebServer
  @ImplementMeasurement{
      PercentageMeasurement CPULoad Minimum 90
  }
}
```
□

*C. Algebraic semantics of complex event operators*

CEDL defines basically three complex event operators, which can be used to combine events into complex events. The operators are summarized in TABLE I.

TABLE I.    COMPLEX EVENT OPERATORS

| Operator | Event | Meaning |
| --- | --- | --- |
| EXISTS(*params*) | point, interval | Parameters should exist. |
| EXISTS(*params*).timewin(τ) | point, interval | Parameters should exist within a time window with length of τ. |
| FOLLOWS(*params*) | point, interval | Parameters follow each other. |
| FOLLOWS_T (*params*; τ) | point, interval | Parameters follow each other regarding a time window with length of τ. |
| CONCURRENT (*params*) | point, interval | Point events appear in the same time. Interval events are happening concurrently. |
| CONCURRENT_T (*params*; τ) | interval only | Interval events are happening concurrently regarding an interval with length of τ. |

TABLE I. shows a comparison of CEDL against two known event algebras. Allen's interval algebra is a calculus for temporal reasoning, applied in embedded software testing. [5] The formalism defines possible relationship operators between two intervals and transition rules in order to propagate relationships among arbitrary number of intervals. The algebra is partially covered by CEDL. The only difference is the refined set of concurrency operators in Allen's algebra, but all these concurrency cases can be expressed by CEDL as well, using the complex event operators and time windows. Besides that, the referred concurrency cases are very rare and uncommon in practice, thus left out intentionally from CEDL.

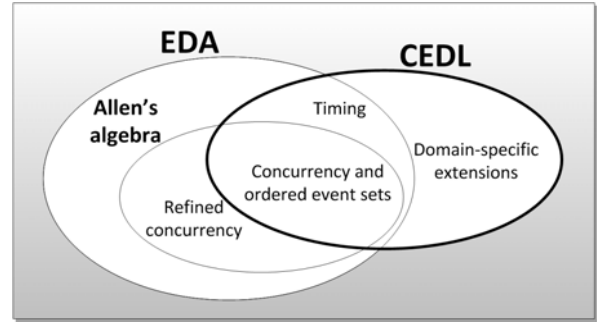

Figure 3. Relations of the algebras discussed in this section.

The Event Detection Algebra, presented by J. Carlson, defines a full-algebraic way to depict event patterns. In his work [6] the author also shows possible implementations. EDA, by its nature has a stronger descriptive power comparing to Allen's algebra and CEDL, since it is built up based on mathematical theorems, not on empirical use-cases. Allen's algebra is fully covered by EDA; however CEDL has the ability to be extended and customized to the targeted domain. Thanks to that, no expensive mathematical apparatus is needed to define detailed patterns.

Eventually, one can draw the following conclusions. CEDL seems to be a feasible tool for the industrial use, since it narrows the language elements of the well-proven Allen's algebra into a reasonable set of operators and extends them with the handling of time dimension, which is a really common use-case in practice. CEDL also has advantages against Carlson's EDA, namely the possibility to extend the core language elements with domain-specific information, thus making the language easy-to-apply for a broad set of different domains.

## IV. UTILIZING DOMAIN INFORMATION

We saw, how domain information is a mandatory part of event patterns. It is major question how domain information can be accessed whilst defining patterns.

*Structural data*, like event sources, is usually stored in an appropriate form. E.g. considering Example 1, event sources are resources in an IT infrastructure and such as these, are usually stored in configuration management databases (CMDB's) and when they are not, it is still easy to collect the required information, for example by a simple discovery over the underlying system. (Figure 1. , Structural data.)

*Measurements and their relevant values (e.g.: thresholds)* raise a much more complicated problem.

In ideal case, every defined event pattern should be associated with one or more high-level goals, since event patterns, as parts of business logic, serve for assuring goals.

In the most simple scenario, during the definition of event patterns, one has to browse the related standards, specifications, SLA's and other high-level documents to determine (a) the event sources, which can successfully capture the phenomenon, described in high-level goals; (b) the measurements needed; (c) the corresponding values of measurements, and assign them manually.

This is not trivial, moreover not feasible in practice, when the underlying system can contain hundreds and thousands of event sources. [7] Thus, this raises exactly the problem of uninsured consistency, as mentioned in Section I.

### A. The semantic knowledge base

Domain information can be derived from the knowledge base in an automated way, by storing domain information not in standard relational databases, but in ontologies, since the possibility of employing reasoning logic in the latter case. By an appropriate structure of the *Semantic Knowledge Base (SKB)*, high-level goals and observable event sources can be associated; therefore the designer of the event patterns does not need to manually collect the suitable event sources, measurements, values, but the reasoning logic can do this.

In addition, reasoning logic can be used to formally prove conformance against goals and standards.

As Figure 4. shows, information contained by the SKB, will be eventually mapped onto the modeling language, thus information about high-level goals (including metadata, such as event source types) will be formalized with the syntax and semantics of the Complex Event Description Language. This implicitly presumes that the knowledge base and the modeling language provide fixed interfaces against each other, since the mapping logic must remain intact, no matter the nature or structure of the targeted domain. We assume furthermore, that the knowledge base we are working with is possibly built up *beforehand* and we have *no influence on the structure* of it (there is no rigorous standard for constructing ontologies).

To fulfill the requirement of fixed interfaces, the SKB is split into two fundamentally different parts. (See Figure 4.)

The *Domain ontology* is the one which actually depicts the domain concepts and relations, in any arbitrary form or structure. Consequently, its structure changes case by case. An adequate example is an ontology depicting business information of a company, or manufacturing and scheduling information of a factory. Dependability concepts and the relations among them can be handled here as well; for instance the taxonomy described in the fundamental paper of the field [8], or by the standards of COBIT [3].

The *Generic ontology* defines the fixed interface, by a fixed set of domain-agnostic concepts, which can be mapped to the language. (See section IV.C.)

At last, Domain ontology is mapped on the Generic ontology. Here, another advantage of ontologies is utilized, namely, the Domain-Generic mapping can be implemented easily by a third ontology, let's call it *Mapping ontology*. This approach is also known as semantic integration. [9]

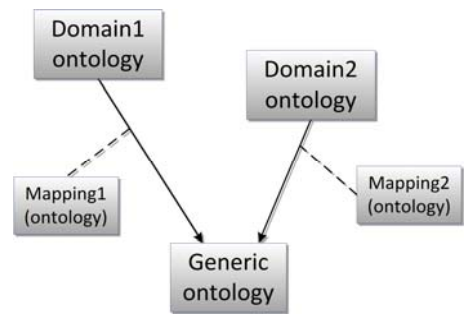

Figure 4. Concept of the Semantic knowledge base of two domains.

### B. Ontology joins

In order to demonstrate how ontology mappings work, the following table defines the General ontology and shows an example Domain ontology; the concepts in one given row can be linked by a mapping.

TABLE II. CONCEPTS OF THE GENERIC ONTOLOGY AND OF AN EXAMPLE DOMAIN ONTOLOGY, FROM THE FIELD OF IT MANAGEMENT.

| Generic ontology concept | "IT" domain example |
|---|---|
| SourceType | *Webserver, DatabaseServer* |
| Quantifier | *load, reliability* |
| Metrics | *m1*(w1: Webserver, load) |
| Range | *r1*(0.9, 1) |
| Qualifier | *critical*(m1, r1) |
| ComplexRelationship | *backup*(w1:Webserver, w2:Webserver) |
| ComplexGroup | *cluster*(w1:Webserver, d1:DatabaseServer) |
| Constraint | *c1*(cluster, at_least_one_Webserver) |

*SourceType* defines an element, which events can be observed of. *Quantifiers* are quantitative indicators. *Metrics* logically link a source of a given SourceType and a Quantifier and can be actually measured, benchmarked. A *Range* is an interval, defined by its starting and ending value. A *Qualifier* defines a qualitative indicator by specifying a Metric for a given Range. Quantifiers and Qualifiers are both usually derived from high-level business goals.

As we need to capture no single source instances, but multiple, combined ones, some grouping concepts are needed in the Generic ontology: *ComplexRelationship* and *ComplexGroup*. The fundamental difference between them is, that the former one is a directed one-to-many or many-to-one association, while the latter one is undirected and formally many-to-many. Of course, ComplexRelationships can exist among the members of a CompleGroup. (E.g.: *backup* relationships in a *cluster* group.)

Finally, by *Constraints*, we have the descriptive power to assign a formalized high-level goal to a ComplexGroup.

The above outlined structure is arbitrary; it was specified based on case studies and real-life industrial problems. Of course, there might be other approaches to provide a general concept-base for CEP.

In this section, I presented, how different domain ontologies of arbitrary structure can be handled and mapped on a general one, thus facilitate a fixed interface towards the General metamodel of the language. Now, we are ready to utilize the domain knowledge to extend the initial crude structure of the modeling language and making it able to capture high-level domain concepts.

### C. Extending the modeling language

After joining the different ontologies, the domain knowledge is mapped onto the language level by fixed rules. This section briefly presents how this is done.

First, we present the Topology metamodel. Every underlying system has a topology consisting of elements, which on the level of the language appear as event sources, but on the level of the actual system, this is a broad structure of physical and logical elements.

Figure 5. summarizes the Topology metamodel, which is close to the one used in the domain of IT infrastructures; and indeed, it is based on CIM (Common Information Metamodel) [10], but it is feasible for other domains as well, thus serves as the base interface of the language for the structural data to be mapped onto.

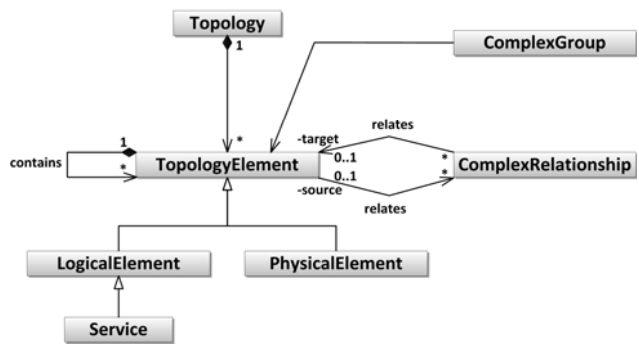

Figure 5. Topology metamodel of underlying systems, based on CIM.

**Example 4** Considering the notations of the example in TABLE II. , so let us be given a source type of name *Webserver*, which is also a *TopologyElement*, more specific a *PhysicalElement*; and let us be given the set of Quantifiers, Qualifiers, Metrics, etc., mentioned there. The mapping logic maps this information onto the CEDL syntax as follows.

From the source type a *SourceType* definition will be generated:
```
SourceType Webserver extends PhysicalElement
```
Next, from the additional information of the *Metrics* and the *Quantifier* the following Event stub will be produced:
```
EventStub LoadWebserverCritical {
  sourceType Webserver
  characteristic Critical
  ScalarMeasurement WebserverCritical In (0.9, 1)
}
```
□

This stub can be implemented by the modeling expert using the CEDL.

We saw how CEDL can be extended by domain information. The fact, that no prior structural standards are need in the *Domain ontology*, should be remarked, since this is the typical case in the real industrial projects.

## V. TOOLING PROTOTYPE

In order to support the process of modeling and to implement the automated code generation feature, I have developed an integrated modeling environment based on up-to-date Eclipse techniques: the textual modeling language is implemented with Xtext 2.0; the code generation logic is implemented with Xtend. [11]

As key features of the IDE, model validation, code completion and assistance and the automated source-code generation (in modeling-time) can be mentioned.

As mentioned already, the modeling language is text-based; prior to implementing, the formalism of UML class diagrams had been tested by. Serious issues had emerged even in very simple cases they underlined the advantage of textual modeling, i.e., it is easier and faster to develop models, than with a graphical tool.

The major drawback of textual modeling emerges when one tries to *explore* a model. It is much easier to comprehend graphical notations. But employing CEP is typical in large-scale environments. Assuming such a vast underlying structure, even the graphical notations of UML fail to present an aggregated, clear model. As a result, the advantage of the usual graphical toolsets is not prevalent in this case, and thus, the textual approach seems to be valid.

## VI. RELATED WORK

In [12] the authors offer a graphical modeling language for defining complex event patterns. Despite the graphical approach, BEMN holds numerous drawbacks compared to CEDL: it doesn't provide any code generation logic in order to be executable on an event processing engine and no executable semantics can be produced from a workflow; it focuses on the dynamics of the event patterns, but doesn't deal with the static structure; because of the previous point, it is not possible to extend the language with full domain-specific knowledge, thus the modeling cannot be aided with this kind of automation; finally, models are built manually, thus in case of a vast problem-space the model could be easily insufficient or incomplete, which defects are not trivial to detect. Nevertheless, the approach could serve as a good starting point for extending CEDL with graphical elements.

Through my research and in this paper as well, I assumed that there exist actually observable events in the underlying structure. An important result of the project CoMiFin [13] is a feasible method for monitoring infrastructures and establishing observable measurement points.

In paper [14], the authors work around a case study of railway vehicle systems where critical dependability attributes need to be assured and deal with the problems of establishing a structure referenced here as the semantic knowledge base. The paper points out, that the semantic knowledge base can be implemented effectively; moreover, its validation is feasible as well.

## VII. CONCLUSIONS

The primary goal of my research work was to develop a model-driven approach for complex event processing that can be automatized to a certain level and to aid this approach with suitable tooling. In this paper, I presented how my modeling language is built up from a core part, which contains domain-agnostic, thus fixed modeling elements; and from a domain-specific part, which enables high-precision event patterns. These two parts effectively complement each other; resulting a lightweight modeling technique and a customizable modeling language.

The ontology-based approach for storing domain-specific knowledge makes this whole approach powerful. High-level goals can be formalized in an arbitrary form, yet the layered architecture of the semantic knowledge base enables extending the language by fixed mappings. Domain information can be maintained in an automated way using measured data, after being preprocessed and analyzed. Reasoning logic can be applied to prove conformance to standards and specifications; or to assign event sources for concrete high-level goals (inductive reasoning).

I compared the algebraic semantics of CEDL to other languages' and pointed out, how it narrows the set of necessary modeling elements to a reasonable level and yet shows the advantage of facilitating high-precision pattern definitions compared to pure algebra-based frameworks (EDA) as a result of the extendibility.

Finally, I have come to the conclusion that CEDL seems to be a *feasible tool* for industrial use, especially recommended in the case of *medium and large scale* underlying systems, where the current tools often fail.

## VIII. FUTURE WORK

My primary goal in the near future is to publish the modeling tool. Prior to that, executing performance tests is the most important task to accomplish. The modeling language also should be tested for other domains, such as e-trading algorithms or proactive systems. [15]

The ontology-based approach enables reasoning logic to be employed while checking whether a given model of events conforms to a given standard of the domain or not.

In a running system, the measured data should be fed back into the semantic knowledge base to keep it up-to-date. This data needs to be preprocessed and analyzed in an intelligent, automated way. I plan this particular aspect to be in the focus of my research work in the next few months.

As mentioned earlier, graphical notation holds its own advantages, especially while describing dynamics of complex events in patterns; thus a domain-specific GUI seems a reasonable extension to the modeling environment and it is currently under planning.

ACKNOWLEDGEMENT

These results could not have been achieved without the constructive ideas and the constant support of my supervisor, *László Gönczy*. I'd like to thank for all his helpful advices, both professional and personal.